\documentclass[amsmath, amssymb, aps, pra,reprint, twocolumn,longbibliography,floatfix,nofootinbib,thesis]{revtex4}
\usepackage[utf8]{inputenc}
\usepackage[urlcolor=blue, colorlinks=true,citecolor=blue]{hyperref}
\usepackage{amsmath}
\usepackage{fancyhdr}
\usepackage{hyperref}
\usepackage{textcomp}
\usepackage{gensymb}
\usepackage{amsmath,amssymb}
\usepackage{tabularx}
\usepackage{zx-calculus}
\usepackage{algorithm}
\usepackage{physics}
\usepackage{algpseudocode}

\begin{document}
\title{Hybrid quantum ResNet for car classification and its hyperparameter optimization}

\author{Asel~Sagingalieva}
\author{Mohammad~Kordzanganeh}
\author{Andrii~Kurkin}
\author{Artem~Melnikov}
\author{Daniil~Kuhmistrov}
\author{Michael~Perelshtein}
\author{Alexey~Melnikov}
\address{Terra Quantum AG, 9000 St.~Gallen, Switzerland}
\thanks{Corresponding author, e-mail: alexey@melnikov.info
\begin{center}
\fbox{
\begin{minipage}{0.46\textwidth}
Please check the published version, which includes all the latest additions and corrections: 
Quantum Mach. Intell. 5:38, 2023, DOI: \href{https://doi.org/10.1007/s42484-023-00123-2}{10.1007/s42484-023-00123-2}
\end{minipage}
}
\end{center}
}

\author{Andrea~Skolik$^{1,3}$}
\author{David~Von~Dollen$^{2,3}$}
\address{$^1$Volkswagen Data:Lab, 80805 Munich, Germany}
\address{$^2$Volkswagen Group of America, Auburn Hills, MI 48326, USA}
\address{$^3$Leiden University, 2333 CA Leiden, The Netherlands}

%\date{\today}

\begin{abstract}
Image recognition is one of the primary applications of machine learning algorithms. Nevertheless, machine learning models used in modern image recognition systems consist of millions of parameters that usually require significant computational time to be adjusted. Moreover, adjustment of model hyperparameters leads to additional overhead. Because of this, new developments in machine learning models and hyperparameter optimization techniques are required. This paper presents a quantum-inspired hyperparameter optimization technique and a hybrid quantum-classical machine learning model for supervised learning. We benchmark our hyperparameter optimization method over standard black-box objective functions and observe performance improvements in the form of reduced expected run times and fitness in response to the growth in the size of the search space. We test our approaches in a car image classification task and demonstrate a full-scale implementation of the hybrid quantum ResNet model with the tensor train hyperparameter optimization. Our tests show a qualitative and quantitative advantage over the corresponding standard classical tabular grid search approach used with a deep neural network ResNet34. A classification accuracy of 0.97 was obtained by the hybrid model after 18 iterations, whereas the classical model achieved an accuracy of 0.92 after 75 iterations.
\end{abstract}

\maketitle

\section*{INTRODUCTION}

The field of quantum computing has seen large leaps in building usable quantum hardware during the past decade. As one of the first vendors, D-Wave provided access to a quantum device that can solve specific types of optimization problems~\cite{johnson2011quantum}. Motivated by this, quantum computing has not only received much attention in the research community, but was also started to be perceived as a valuable technology in industry. Volkswagen published a pioneering result on using the D-Wave quantum annealer to optimize traffic flow in 2017~\cite{neukart2017traffic}, which prompted a number of works by other automotive companies~\cite{mehta2019quantum,ohzeki2019control,yarkoni2021multi}. Since then, quantum annealing has been applied in a number of industry-related problems like chemistry~\cite{streif2019solving,xia2017electronic}, aviation~\cite{stollenwerk2019quantum}, logistics~\cite{feld2019hybrid} and finance~\cite{grant2021benchmarking}. Aside from quantum annealing, gate-based quantum devices have gained increased popularity, not least after the first demonstration of a quantum device outperforming its classical counterparts~\cite{arute2019quantum}. A number of industry-motivated works have since been published in the three main application areas that are currently of interest for gate-based quantum computing: optimization~\cite{streif2021beating,streif2020training,amaro2021case,dalyac2021qualifying,luckow2021quantum}, quantum chemistry and simulation~\cite{arute2020hartree,malone2021towards}, and machine learning~\cite{melnikov2023quantum,rudolph2020generation,skolik2021layerwise,Skolik2022quantumagentsingym,peters2021machine,alcazar2020classical,boston-housing,sagingalieva2022hybrid,kordzanganeh2022exponentially}. Research in the industrial context has been largely motivated by noisy intermediate-scale quantum (NISQ) devices~\cite{kordzanganeh2022benchmarking} -- early quantum devices with a small number of qubits and no error correction. In this regime, variational quantum algorithms (VQAs) have been identified as the most promising candidate for near-term advantage due to their robustness to noise~\cite{cerezo2021variational}. In a VQA, a parametrized quantum circuit (PQC) is optimized by a classical outer loop to solve a specific task like finding the ground state of a given Hamiltonian or classifying data based on given input features. As qubit numbers are expected to stay relatively low within the next years, hybrid alternatives to models realized purely by PQCs have been explored~\cite{zhang2021variational,mari2020transfer,zhao2019qdnn,dou2021unsupervised,sebastianelli2021circuit,pramanik2021quantum,boston-housing,sagingalieva2022hybrid}. In these works, a quantum model is combined with a classical model and optimized end-to-end to solve a specific task. In the context of machine learning, this means that a PQC and neural network (NN) are trained together as one model, where the NN can be placed either before or after the PQC in the chain of execution. When the NN comes first, it can act as a dimensionality reduction technique for the quantum model, which can then be implemented with relatively few qubits.

In this work, we use a hybrid quantum-classical model to perform image classification on a subset of the Stanford Cars data set~\cite{KrauseStarkDengFei-Fei_3DRR2013}. Image classification is an ubiquitous problem in the automotive industry, and can be used for tasks like sorting out parts with defects. Supervised learning algorithms for classification have also been extensively studied in quantum literature \cite{havlivcek2019supervised,schuld2019quantum,schuld2020circuit,rebentrost2014quantum}, and it has been proven that there exist specific learning tasks based on the discrete logarithm problem where a separation between quantum and classical learners exists for classification~\cite{liu2021rigorous}. While the separation in~\cite{liu2021rigorous} is based on Shor's algorithm and therefore not expected to transfer to realistic learning tasks as the car classification mentioned above, it motivates further experimental study of quantum-enhanced models for classification on real-world data sets.

In combining PQCs and classical NNs into hybrid quantum-classical models, we encounter a challenge in searching hyperparameter configurations that produce performance gains in terms of model accuracy and training. Hyperparameters can be considered values that are set for the model and do not change during the training regime, and may include variables such as learning rate, decay rates, choice of optimizer for the model, number of qubits or layer sizes. Often in practice, these parameters are selected by experts based upon some a priori knowledge and trial-and-error. This limits the search space, but in turn can lead to producing a suboptimal model configuration.

Hyperparameter optimization is the process of automating the search for the best set of hyperparameters, reducing the need for expert knowledge in hyperparameter configurations for models, with an increase in computation required to evaluate configurations of models in search of an optimum. In the 1990s, researchers reported performance gains leveraging a wrapper method, which tuned parameters for specific models and data sets using best-first search and cross validation~\cite{KOHAVI1995304}. In more recent years, researchers have proposed search algorithms using bandits~\cite{Li2017}, which leverage early stopping methods. Successive Halving algorithms such as the one introduced in~\cite{pmlr-v28-karnin13} and the parallelized version introduced in~\cite{Li2018} allocate more resources to more promising configurations. Sequential model-based optimization leverages Bayesian optimization with an aggressive dual racing mechanism, and also has shown performance improvements for hyperparameter optimization~\cite{Hutter2011, Lindauer_Hutter_2018}. Evolutionary and population-based heuristics for black-box optimization have also achieved state-of-the-art results when applied to hyperparameter optimization in numerous competitions for black-box optimization \cite{vermetten2020sequential,back1996,awad2020squirrel}. In recent years, a whole field has formed around automating the process of finding optimal hyperparameters for machine learning models, with some prime examples being neural architecture search~\cite{elsken2019neural} and automated machine learning (AutoML)~\cite{hutter2019automated}. Automating the search of hyperparameters in a quantum machine learning (QML) context has also started to attract attention, and the authors of~\cite{berganza2022towards} have explored the first version of AutoQML.

Our contribution in this paper is not only to examine the performance gains of hybrid quantum-classical models vs. purely classical, but also to investigate whether quantum-enhanced or quantum-inspired methods may offer an advantage in automating the search over the configuration space of the models. We show a reduction in computational complexity in regard to expected run times and evaluations for various configurations of models, the high cost of which motivate this investigation. We investigate using the tensor train decomposition for searching the hyperparameter space of the HQNN framed as a global optimization problem as in~\cite{globalOptimizationTensorTrains}. This method has been successful in optimizing models of social networks in~\cite{sergey2019tensor}, and as a method of compression for deep neural networks~\cite{WANG2021320}.

\section*{RESULTS}

\subsection{Hyperparameter Optimization}
\begin{figure*}[ht!]
    \centering
    \includegraphics[width=1\linewidth]{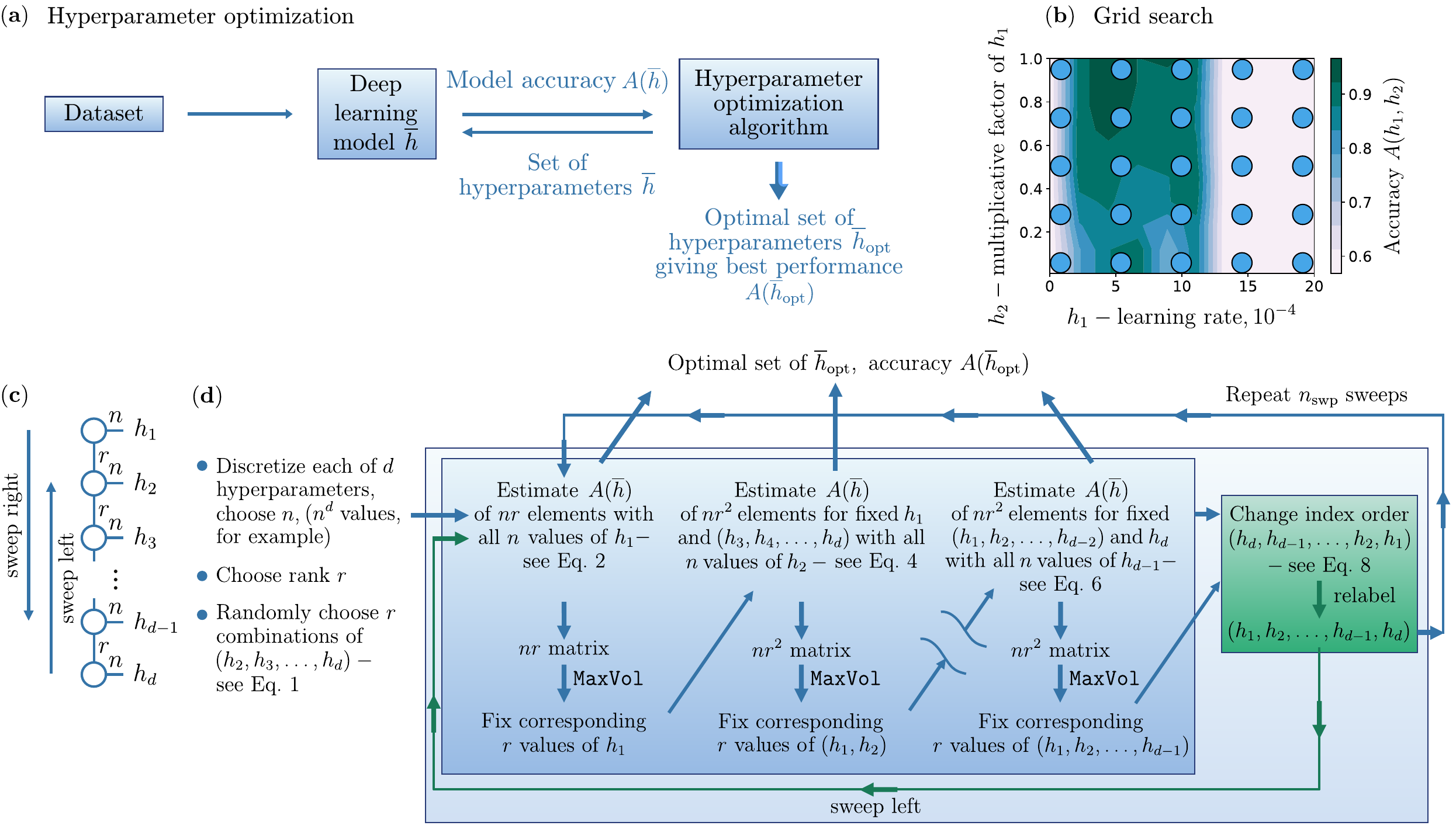}
    \caption{The hyperparameter optimization problem description (a). The tabular methods for hyperparameter optimization: the grid search algorithm (b) and the tensor train algorithm (c-d).}
    \label{fig:HPO}
\end{figure*}

The problem of hyperparameter optimization (HPO) is described schematically in Fig.~\ref{fig:HPO}(a). Given a certain data set and a machine learning (ML) model, the learning model demonstrates an accuracy $A(\bar{h})$ which depends on the hyperparameters $\bar{h}$. To achieve the best possible model accuracy, one has to optimize the hyperparameters. To perform the HPO, an unknown black-box function $A(\bar{h})$ has to be explored. The exploration is an iterative process, where at each iteration the HPO algorithm provides a set of hyperparameters $\bar{h}$ and receives the corresponding model accuracy $A(\bar{h})$. As a result of this iterative process, the HPO algorithm outputs the best achieved performance $A(\bar{h}_\mathrm{opt})$ with the corresponding hyperparameters $\bar{h}_\mathrm{opt}$.

The HPO could be organized in different ways. One of the standard methods for HPO is a tabular method of grid search (GS), also known as a parameter sweep (Fig.~\ref{fig:HPO}(b)). To illustrate how a grid search works, we have chosen two hyperparameters: the learning rate ($h_1$) and the multiplicative factor of learning rate ($h_2$). They are plotted along the $x$-axis and the $y$-axis, respectively. The color on the contour shows the accuracy of the model $A(h_1,h_2)$ with two given hyperparameters changing from light pink (the lowest accuracy) to dark green (the highest accuracy). In the GS method, the hyperparameter values are discretized, which results in a grid of values shown as big dots. The GS algorithm goes through all the values from this grid with the goal of finding the maximum accuracy. As one can see in this figure, there are only three points at which this method can find a high accuracy with 25 iterations (shown as 25 points in Fig.~\ref{fig:HPO}(b)). This example shows that there could be a better tabular HPO in terms of the best achievable accuracy and the number of iterations used.

\subsection{Tensor train approach to hyperparameter optimization}

Here, we propose a quantum-inspired approach to hyperparameter optimization based on the tensor train (TT) programming. The TT approach was initially introduced in the context of quantum many-body system analysis, e.g., for finding a ground state with minimal energy of multi-particle Hamiltonians via Density Matrix Renormalization Groups (DMRG)~\cite{DMRG}. In this approach, the ground state is represented in the TT format, often referred to as the Matrix Product State in physics~\cite{MPS}. We employ the TT representation (shown in Fig.~\ref{fig:HPO}(c)) in another way here, and use it for the hyperparameter optimization. As one can see in Fig.~\ref{fig:HPO}(c), the TT is represented as a multiplication of tensors, where an individual tensor is shown as a circle with the number of ``legs'' that corresponds to the rank of the tensor. $h_1$ and $h_d$ circles are the matrices of $n\times r$ dimension, and $\{h_i\}_{i={2}}^{i={d-1}}$ is a rank 3 tensor of dimensions $n \times r^2$. The two arrows in the Fig.~\ref{fig:HPO}(c) illustrate sweeps right and left along with the TT. This refers to the algorithm described below. Leveraging the locality of the problem, i.e., a small correlation between hyperparameters, we perform the black-box optimization based on the cross-approximation technique applied for tensors~\cite{tt_cross, tt_min}.

Similar to the previously discussed GS method, we discretize the hyperparameter space with TT optimization (TTO) and then consider a tensor composed of scores that can be estimated by running an ML model with a corresponding set of hyperparameters. However, compared to GS, the TT method is dynamic, which means that the next set of evaluating points in the hyperparameter space is chosen based on the knowledge accumulated during all previous evaluations. With TTO we will not estimate all the scores $A(\bar{h})$ available to the model. Instead of this, we will approximate $A(\bar{h})$ via TT, referring to a limited number of tensor elements using the cross-approximation method~\cite{tt_cross}. During the process, new sets of hyperparameters for which the model needs to be evaluated are determined using the {\tt MaxVol} routine~\cite{maxvol}. The {\tt MaxVol} routine is an algorithm that finds an $r \times r$ submatrix of maximum volume, i.e., a square matrix with a maximum determinant module in an $n \times r$ matrix. 

Hyperparameters are changed in an iterative process, in which one is likely to find a better accuracy $A(\bar{h})$ after each iteration, and thus find a good set of hyperparameters. Notably, the TTO algorithm requires an estimate of $\mathcal{O}(d n r^2)$ elements and $\mathcal{O}(d n r^3)$ of calculations, where $d$ is the number of hyperparameters, $n$ is a number of discretization points, and $r$ is a fixed rank. If one compares it with the GS algorithm, which requires estimation of $\mathcal{O}(n^d)$ elements, one is expected to observe practical advantages, especially with a large number of hyperparameters.

\begin{algorithm}
\caption{Tensor Train Optimization}\label{alg:TTO}
\begin{algorithmic}[1]
\State Accuracy $A(\bar{h}_\mathrm{opt}) = 0$
\State $i_{\mathrm{swp}} = 1$
\State $\mathrm{Core} = 1$
\State $j_{\mathrm{swp}} = 1$
\State Discretize each of $d$ hyperparameters with $n$ points
\State Randomly choose $r$ combinations of $(h_2, h_3, \ldots , h_d)$
\While{$i_{\mathrm{swp}} \leq n_{\mathrm{swp}}$}
\While{$j_{\mathrm{swp}} \le 2$}
\While{$\mathrm{Core} < d$}
\If{$\mathrm{Core} == 1$}
\State Estimate $A(\bar{h})$ of $nr$ elements with all $n$ values of $h_1$
\If{$A(\bar{h}_\mathrm{opt}) < A(\bar{h})$} \State $A(\bar{h}_\mathrm{opt}) = A(\bar{h})$
\EndIf
\State {\tt MaxVol}
\State Fix corresponding $r$ values of $h_1$
\Else
\State Estimate $A(\bar{h})$ of $nr^2$ elements for fixed $(h_{1}, \ldots, h_{\mathrm{Core}-1})$ with all $n$ values of $h_{\mathrm{Core}}$
\If{$A(\bar{h}_\mathrm{opt}) < A(\bar{h})$} 
\State $A(\bar{h}_\mathrm{opt}) = A(\bar{h})$
\EndIf
\State {\tt MaxVol}
\State Fix corresponding $r$ values of $(h_1, \ldots, h_{\mathrm{Core}})$
\EndIf
\State $\mathrm{Core} = \mathrm{Core} + 1$
\EndWhile
\State Change index order $(h_d, \ldots, h_1)$
\State Relabel  $(h_1, \ldots, h_d)$
\State $\mathrm{Core} = 1$
\State $j_{\mathrm{swp}} = j_{\mathrm{swp}} + 1$
\EndWhile
\State $j_{\mathrm{swp}} = 1$
\State $i_{\mathrm{swp}} = i_{\mathrm{swp}}+1$
\EndWhile
\end{algorithmic}
\end{algorithm}

The TTO algorithm for the HPO is presented as the Algorithm~\ref{alg:TTO} pseudocode that also corresponds to Fig.~\ref{fig:HPO}(d). The TTO algorithm can be described with $9$ steps:
\begin{enumerate}
    \item Suppose each of $d$ hyperparameters is defined on some interval $h_i \in [h_i^\mathrm{min}, h_i^\mathrm{max}]$, where $i \in [1, d]$.
    One first discretizes each of $d$ hyperparameters by defining $n$ points 
    \begin{displaymath}
     {\{ h_i(1), h_i(2), \ldots , h_i(n)\}}_{i=1}^{i=d}.   
    \end{displaymath}
    \item Then, we need to choose the rank $r$. This choice is a trade-off between computational time and accuracy, which respectively require a small and a large rank.
    \item $r$ combinations of 
    \begin{equation}
    {\{h_2^1(j), h_3^1(j), \ldots , h_d^1(j)\}}_{j=1}^{j=r}
    \end{equation}
    are chosen.
    \item In the next three steps, we implement an iterative process called the ``sweep right''. The first step of this iterative process is related to the first TT core evaluation:
            \begin{itemize}
            \item  The accuracy of $nr$ elements is estimated with all $n$ values of the first hyperparameter ${\{h_1(i_1)\}}_{i_1=1}^{i_1=n}$ and for the $r$ combinations of ${\{h_2^{1}(j), h_3^{1}(j), \ldots , h_d^{1}(j)\}}_{j=1}^{j=r}$:
            \begin{equation}
            \begin{split}
            {\{A(h_1(i_1), h_2^1(j), h_3^1(j),  \ldots,} \\  
            {h_d^1(j))\}}_{j=1, i_1=1}^{j=r, i_1=n}.         
            \end{split}
            \end{equation}
            \item In this matrix of size $n \times r$ we search for a submatrix with maximum determinant module:
            \begin{equation}
            \quad \quad {\{A(h_1^1(i_1), h_2^1(j), h_3^1(j), h_d^1(j))\}}_{j=1, i_1=1}^{j=r, i_1=r}.
            \end{equation}
            The corresponding $r$ values of the first hyperparameter are fixed $\{h_1^1(i_1)\}_{i_1=1}^{i_1=r}$.
            \end{itemize}
    \item The next step of this iterative process is related to the second TT core evaluation:
            \begin{itemize}
            \item We fix $r$ values $\{h_1^1(i_1)\}_{i_1=1}^{i_1=r}$ of the previous step as well as $r$ combinations ${\{h_3^1(j), h_4^1(j), \ldots , h_d^1(j)\}}_{j=1}^{j=r}$ of the third step. We, then, estimate the accuracy of the $nr^2$ elements with all $n$ values of the second hyperparameter ${\{h_2(i_2)\}}_{i_2=1}^{i_2=n}$:
            \begin{equation}
            \begin{split}
                {\{A(h_1^1(i_1), h_2(i_2), h_3^1(j), \ldots ,} \\ {h_d^1(j))\}}_{j=1, i_1=1, i_2=1}^{j=r, i_1=r, i_2=n}`
            \end{split}                
            \end{equation}
            \item Again, in this matrix of size $nr \times r$ we search for a submatrix with the maximum determinant module:
            \begin{equation}
            \begin{split}
                {\{A((h_1^2(k), h_2^2(k)), h_3^1(j), \ldots,} \\ {h_d^1(j))\}}_{j=1, k=1}^{j=r, k=r}
            \end{split}
            \end{equation}

            $r$ combinations ${\{(h_1^2(k), h_2^2(k))\}}_{k=1}^{k=r}$ of the first and the second hyperparameters are fixed.
    \end{itemize}
    \item The $d-1$ TT core evaluation:
    \begin{itemize}
     \item We fix $r$ combinations ${\{(h_1^{d-2}(k), h_2^{d-2}(k), \ldots , h_{d-2}^{d-2}(k))\}}_{k=1}^{k=r}$ of the $d-2$ TT core as well as $r$ combinations ${\{h_d^1(j)\}}_{j=1}^{j=r}$ of the third step. We, then, estimate the accuracy of the $nr^2$ elements with all $n$ values of the ${\{h_{d-1}(i_d)\}}_{i_d=1}^{i_d=n}$:
    \begin{equation}
    \begin{split}
        {\{A((h_1^{d-2}(k), \ldots , h_{d-2}^{d-2}(k)), }\\ 
        {h_{d-1}(i_{d-1}), h_d^1(j))\}}_{k=1,i_{d-1}=1, j=1}^{k=r, i_{d-1}=n, j=r}  
    \end{split}
    \end{equation}
    \item Again, in this matrix of size $nr \times r$ we search for a submatrix with the maximum determinant module:
        \begin{equation}
        \begin{split}
          {\{A((h_1^{d-1}(k), h_2^{d-1}(k), \ldots ,} \\ 
          {h_{d-1}^{d-1}(k)), h_d^1(j))\}}_{k=1, j=1}^{k=r, j=r}
        \end{split}
        \end{equation}
    
    $r$ combinations of ${\{(h_1^{d-1}(k), h_2^{d-1}(k), \ldots, h_{d-1}^{d-1}(k)\}}_{k=1}^{k=r}$ hyperparameters are fixed.
    \end{itemize}
    The end of one ``sweep right'' is reached.
    \item Similar to step 3, we have $r$ combinations of hyperparameters, but they are not random anymore. We next perform a similar procedure in the reverse direction (from the last hyperparameter to the first). The process is called the ``sweep left''.
    
   One first changes the index order:
    \begin{equation}
    \quad {\{(h_1^{d-1}(k), h_2^{d-1}(k), \ldots, h_{d-1}^{d-1}(k)\}}_{k=1}^{k=r} \Longrightarrow \mathrm{relabel}\nonumber
    \end{equation}
    \begin{equation}
        {\{(h_{d-1}^{d-1}(k), h_{d-2}^{d-1}(k), \ldots, h_{2}^{d-1}(k)\}}_{j=1}^{j=r}
    \end{equation}

    And then, continues from the fourth step of the TTO algorithm.

    \item A combination of the ``sweep right'' and the ``sweep left'' is a full sweep. We do $n_\mathrm{swp}$ full sweeps in this algorithm.
    \item During all the iterations, we record it if we estimate a new maximum score. 
\end{enumerate}

\subsection{Benchmarking HPO Methods}

\begin{figure}[ht!]
    \centering
    \includegraphics[width=1\linewidth]{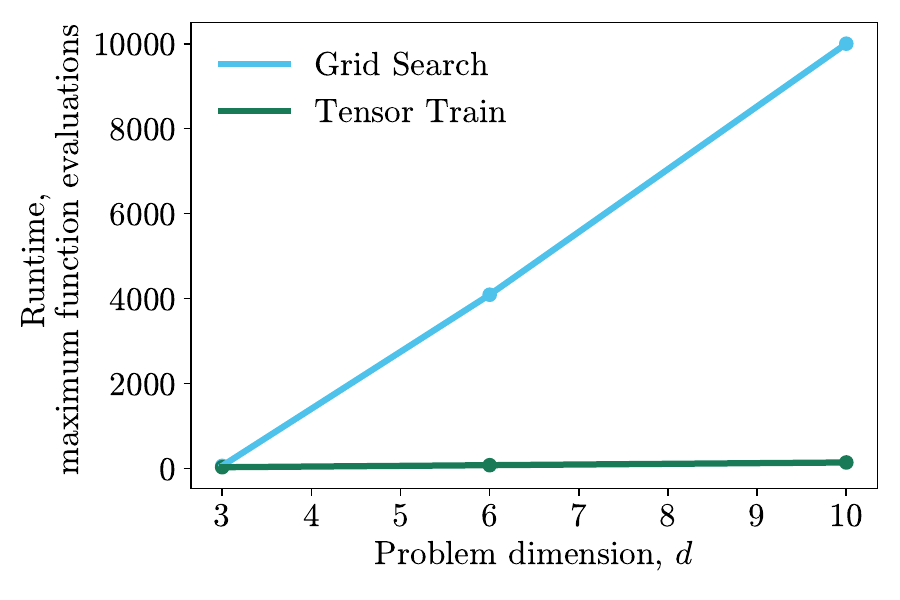}
    \caption{Tensor Train (TT) and Grid Search (GS): Expected Runtime in maximum objective function evaluations vs. growth of problem dimension $d$.}
    \label{fig:Benchmark}
\end{figure}

In order to ascertain the solution quality in our proposed method for hyperparameter optimization, we tested over three black-box objective functions. These functions included the Schwefel, Fletcher-Powell, and Vincent functions from the {\tt optproblems} Python library~\cite{OptProblems}. We ran 100 randomly initialized trails and recorded average fitness and maximum number of function evaluations in response to the change in the problem size $d$ for each objective function. We compared grid search (GS) and tensor train (TT) - both tabular methods for hyperparameter optimization. For tensor train and grid search, we partitioned the hyperparameter ranges with 4 discrete points per hyperparameter. For tensor train we set the rank parameter $r=2$. 

\begin{table}
\begin{tabular}{|l|l|l|l|}
    \hline
    \multicolumn{2}{|c|}{\textbf{Schwefel }} & & \\
    \hline
    \textbf{HPO Method} & \textbf{Average Fitness} & \textbf{$d$} & \textit{ER}\\
    \hline
    \textbf{TT} &  \textbf{-541.76} & \textbf{3} & \textbf{32}\\
    GS & -541.76 & 3 & 64 \\
    \hline
    \textbf{TT} &  \textbf{-1083.53 }& \textbf{6} & \textbf{80} \\
    GS & -1083.53 & 6  & 4092\\
    \hline
    \textbf{TT} &  \textbf{-1805.89} & \textbf{10} & \textbf{144} \\
    GS & -1805.89 & 10 & 10000\\
    \hline
    \multicolumn{2}{|c|}{\textbf{Fletcher-Powell }} &  &\\
    \hline
    \textbf{HPO Method}& \textbf{Average Fitness} & \textbf{$d$}& \textit{ER} \\
    \hline
    TT &  5136.64 & 3 &32 \\
    \textbf{GS} & \textbf{4113.78} & \textbf{3} & \textbf{64}\\
    \hline
    TT &  23954.5 & 6 & 80\\
    \textbf{GS} & \textbf{14295.2} & \textbf{6} & \textbf{4092}\\
    \hline
    TT &  78101.4 &  10 & 144\\
    \textbf{GS} &  \textbf{36890.11} & \textbf{10} & \textbf{10000}\\
    \hline
    \multicolumn{2}{|c|}{\textbf{Vincent }} & & \\
    \hline
    \textbf{HPO Method} & \textbf{Average Fitness} & \textbf{$d$} & \textit{ER} \\
    \hline
    TT & -0.232 & 3 & 32\\
   \textbf{GS} & \textbf{-0.243} & \textbf{3} & \textbf{64}\\
    \hline
    TT &  -0.242 & 6 & 80\\
    \textbf{GS} &  \textbf{-0.243} & \textbf{6} &\textbf{4092}\\
    \hline
    TT &  -0.241 & 10 &144\\
    \textbf{GS} & \textbf{-0.243} & \textbf{10} &\textbf{10000}\\
    \hline
\end{tabular}
\caption{\label{tab:HPO Benchmarks}Table of results comparing HPO methods for Schwefel, Fletcher-Powell, and Vincent objective functions. Average fitness values and Expected Runtimes (\textit{ER}) in maximum function evaluations were calculated over 100 runs for varying sizes of problem dimension $d$ (lower is better). Methods obtaining the best average fitness are highlighted in bold, with ties broken by lower \textit{ER}.}

\end{table}
\begin{figure*}[ht!]
    \centering
    \includegraphics[width=1\linewidth]{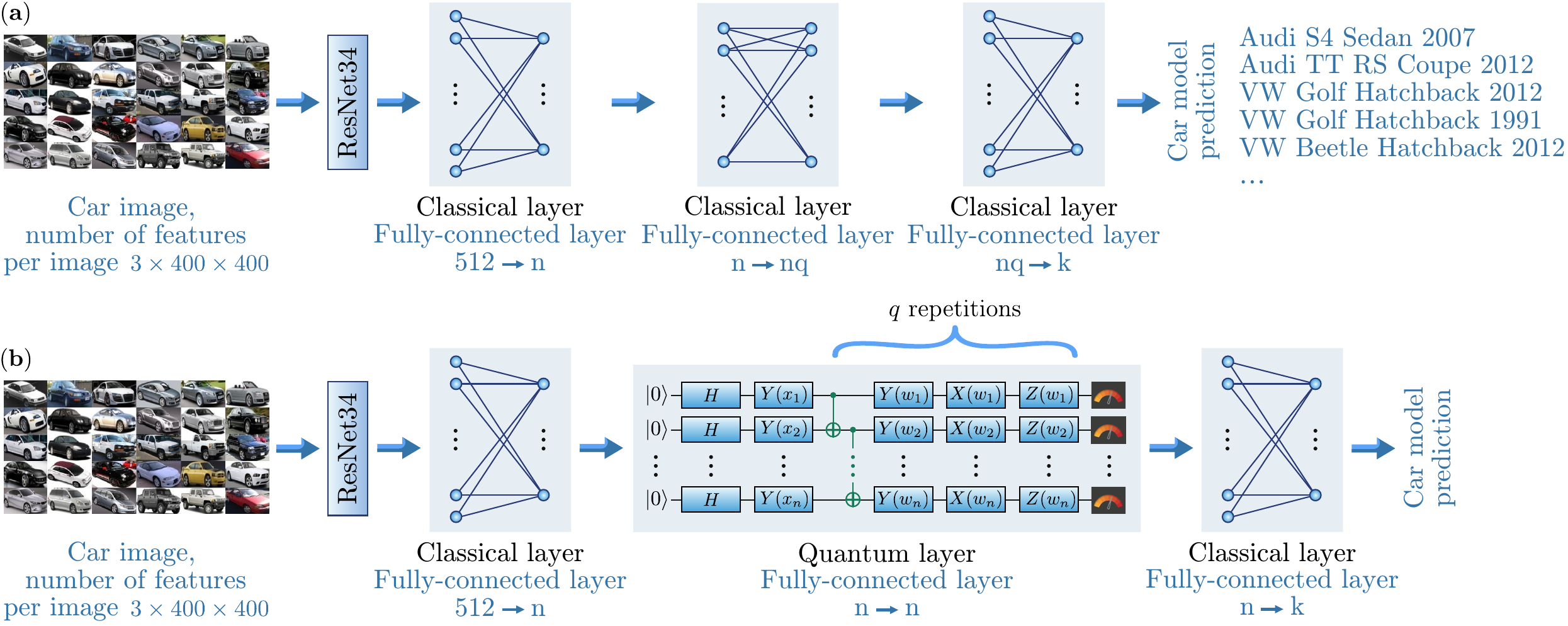}
    \caption{Classical (a) and Hybrid (b) quantum neural network architectures.}
    \label{fig:HQNN}
\end{figure*}

\subsection{Car Classification with Hybrid Quantum Neural Networks}\label{sec:HQNN}

Computer vision and classification systems are ubiquitous within the mobility and automotive industries. In this article, we investigate the car classification problem using the Car data set~\cite{Cars} provided by Stanford CS Department. Examples of cars in the data set are shown in Fig.~\ref{fig:HQNN}. The Stanford Cars data set contains 16,185 images of 196 classes of cars. The data is split into 8,144 training images and 8,041 testing images. The classes are typically at the combination of Make, Model, Year, e.g., Volkswagen Golf Hatchback 1991 or Volkswagen Beetle Hatchback 2012. Since the images in this data set have different sizes, we resized all images to 400 by 400 pixels. In addition, we apply random rotations by maximum 15\degree, random horizontal flips, and normalization to the training data. For testing data, only normalization has been applied.

We use transfer learning to solve the car classification problem. Transfer learning is a powerful method for training neural networks in which experience in solving one problem helps in solving another problem~\cite{Transfer_Learning}. In our case, the ResNet (Residual Neural Network)~\cite{ResNet} is pretrained on the ImageNet data set~\cite{ImageNet}, and is used as a base model. One can fix the weights of the base model, but if the base model is not flexible enough, one can ``unfreeze'' certain layers and make it trainable. Training deep networks is challenging due to the vanishing gradient problem, but ResNet solves this problem with so-called residual blocks: inputs are passed to the next layer in the residual block. In this way, deeper layers can see information about the input data. ResNet has established itself as a robust network architecture for solving image classification problems. We dowloaded ResNet34 via PyTorch~\cite{PyTorch}, where the number after the model name, 34, indicates the number of layers in the network.

As shown in the Fig.~\ref{fig:HQNN}(a), in the classical network after ResNet34 we add three fully-connected layers. Each output neuron corresponds to a particular class of the classification problem, e.g., Volkswagen Golf Hatchback 1991 or Volkswagen Beetle Hatchback 2012. The output neuron with the largest value determines the output class. Since the output from the ResNet34 is composed of 512 features, the first fully-connected layer consists of 512 input neurons and a bias neuron and $n$ output features. The second fully-connected layer connects $n$ input neurons and a bias neuron with $nq$ output features. The value of $n$ and $q$ can vary, thus changing the number of weights in the classical network. Since the network classifies $k$ classes in the general case, the third fully-connected layer takes $nq$ neurons and a bias neuron as input and feeds $k$ neurons as output.

In the hybrid analog as shown in Fig.~\ref{fig:HQNN}(b) we replace the second fully-connected layer with a quantum one. It is worth noting that the number of qubits used for the efficient operation of the model is initially unknown. In the quantum layer, the Hadamard transform is applied to each qubit, then the input data is encoded into the angles of rotation along the $y$-axis. The variational layer consists of the application of the CNOT gate and rotation along $x$, $y$, $z$-axes. The number of variational layers can vary. Accordingly, the number of weights in the hybrid network can also change. The measurement is made in the $X$-basis. For each qubit, the local expectation value of the $X$ operator is measured. This produces a classical output vector, suitable for additional post-processing. Since the optimal number of variational layers ($q$, depth of quantum circuit) and the optimal number of qubits $n$ are not known in advance, we choose these values as hyperparameters. A thorough analysis of the quantum circuit for $n=2$  is given in the Appendix, where three approaches are employed to measure the efficiency, trainability, and the expressivity of this quantum model. 

We use the cross-entropy as a loss function 
\begin{equation}
    l = -\sum_{c=1}^{k}{y_c \log p_c}
\end{equation}
where $p_c$ is the prediction probability, $y_c$ is 0 or 1, determining respectively if the image belongs to the prediction class, and $k$ is the number of classes. We use the Adam optimizer~\cite{Adam,kingma2014adam} and reduce the learning rate after several epochs. There is no one-size-fits-all rule of how to choose a learning rate. Moreover, in most cases, dynamic control of the learning rate of a neural network can significantly improve the efficiency of the backpropagation algorithm. For these reasons, we choose the initial learning rate, the period of learning rate decay, and the multiplicative factor of the learning rate decay as hyperparameters. In total, together with number of variational layers and number of qubits, we optimize five hyperparameters presented in Table~\ref{tab:Hyperparameters} to improve the accuracy of solving the problem of car classification.

\begin{table}
\begin{tabular}{|p{3.2cm}|p{0.9cm}|p{1cm}|p{1.1cm}|p{1.4cm}|}
    \hline
    \textbf{Hyperparameter} & \textbf{Label} & \textbf{Range} & \textbf{Hybrid HPO values} & \textbf{Classical HPO values}
    \\
    \hline
    number of qubits, number of neurons & $n$ & $4-16$ & $13$ & $5$\\
    
    \hline
    depth of quantum circuit & $q$ & $1-5$ & $4$ & $\times$\\
    \hline
    number of neurons & $nq$ & $4-80$ & $\times$ & $80$\\
    
    \hline
    initial learning rate &  $\alpha_{0}$ & $1-10 \times 10^{-4}$ & $5 \times 10^{-4}$ & $5 \times 10^{-4}$\\
    
    \hline
    step of learning rate & $\alpha_{\delta}$ & $1-8$ & $8$ & $5$\\
    \hline
    multiplicative factor of learning rate decay & $\alpha_{r}$ & $0.1-0.2$ & $0.1$ & $0.2$\\
    
    \hline
\end{tabular}
\caption{\label{tab:Hyperparameters}The table shows which hyperparameters are being optimized, their labels, limits of change, and the best values found during HPO.}
\end{table}
    
\begin{figure*}[ht!]
    \centering
    \includegraphics[width=1\linewidth]{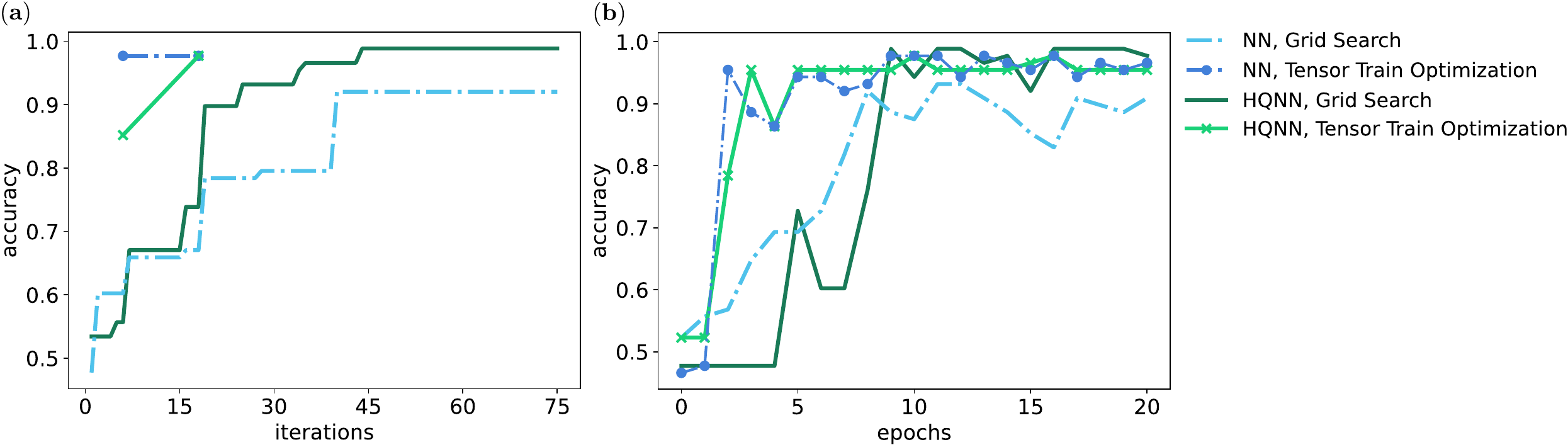}
    \caption{(a) Dependence of accuracy on the number of iterations HPO. TTO for the hybrid model found a set of hyperparameters that gives an accuracy of 0.852 after 6 iterations, 0.977 after 18 iterations, for the classical model found 0.977 after 6 iterations. Grid search for the hybrid model found a set of hyperparameters that gives an accuracy of 0.989 after 75 iterations, for the classical model found 0.920 after 75 iterations. (b) Dependence of accuracy on the number of epochs with the found optimal set of hyperparameters.}
    \label{fig:Results}
\end{figure*}

\subsection{Simulation Results}

We next perform a simulation of the hybrid quantum residual neural network described in the previous section. The simulation is compared to its classical analog, the residual neural network, in a test car classification task. Because of the limited number of qubits available and computational time constraints, we used a classification between two classes, Volkswagen Golf Hatchback 1991 and Volkswagen Beetle Hatchback 2012, to compare the classical and hybrid networks fairly. In total, we used 88 testing images and 89 training images. Both the hybrid quantum HQNN model, and the classical NN model, were used together with the GS and TTO methods for hyperparameter optimization. All machine learning simulations were carried out in the QMware cloud, on which the classical part was implemented with the PyTorch framework, and the quantum part was implemented with the $<$basiq$>$ SDK~\cite{QMWare,boston-housing,kordzanganeh2022benchmarking}. The results of the simulations are shown in Fig.~\ref{fig:Results}.

Fig.~\ref{fig:Results}(a) shows the dependence of accuracy on the number of HPO iterations on the test data, where one iteration of HPO is one run of the model. Green color shows the dependence of accuracy on the number of iterations for the HQNN, blue color shows for the classical NN. As one can see from Fig.~\ref{fig:Results}(a), TTO works more efficiently than GS and in fewer iterations finds hyperparameters that give an accuracy above 0.9. HQNN with TTO (marked with green crosses) finds a set of hyperparameters that yields 97.7\% accuracy over 18 iterations. As for the GS (marked solid green line), it took 44 iterations to pass the threshold of 98\% accuracy.

TTO finds in 6 iterations a set of hyperparameters for the classical NN, which gives an accuracy of 97.7\%, which is the same as the accuracy given by the set of hyperparameters for the HQNN that found in 18 iterations. As for the GS, it is clear that the optimization for the HQNN works more efficiently than for the classical one. And the optimization of the HQNN requires fewer iterations to achieve higher accuracy compared to the optimization of the classical NN. A possible reason is that a quantum layer with a relatively large number of qubits and a greater depth works better than its classical counterpart.

\begin{figure}[ht!]
    \centering
    \includegraphics[width=0.6\linewidth]{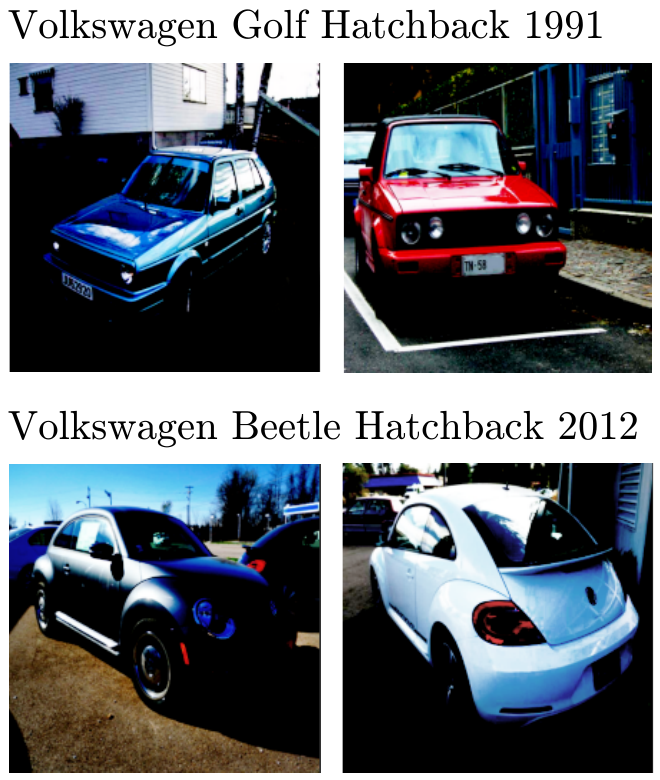}
    \caption{Examples of test car images that were correctly classified by the hybrid quantum residual neural network.}
    \label{fig:Result-Cars}
\end{figure}

 The best values found during HPO are displayed in Table~\ref{tab:Hyperparameters}. The quantum circuit corresponding to the optimal set of hyperparameters has 52 variational parameters, leading to a total of 6749 weights in the HQNN. In the classical NN there are 9730 weights. Therefore, there are significantly fewer weights in a HQNN compared to a classical NN. Nevertheless, as can be seen from the Fig.~\ref{fig:Results}(b), the HQNN, with the hyperparameters found using the GS, reaches the highest overall accuracy (98.9\%). The Fig.~\ref{fig:Result-Cars} shows examples of car images that were classified correctly by the HQNN model. The HQNN with an optimized set of hyperparameters achieved an accuracy of 0.989.

\section*{DISCUSSION}

We introduced two new ML developments to image recognition. First, we presented a quantum-inspired method of tensor train decomposition for choosing ML model hyperparameters. This decomposition enabled us to optimize hyperparameters similar to other tabular search methods, e.g., grid search, but required only $\mathcal{O}(d n r^2)$ hyperparameter choices instead of $\mathcal{O}(n^d)$ in the grid search method. We verified this method over various black box functions and found that the tensor train method achieved comparable results in average fitness, with a reduced expected run time for most of the test functions compared to grid search. This indicates that this method may be useful for high dimensional hyperparameter searches for expensive black-box functions. Future work could investigate using this method in combination with local search heuristic, where the tensor train optimizer performs a sweep over a larger search space within a budget, and seeds another optimization routine for a local search around this region. This method could also be applied to the $B/n$ problem for successive halving algorithm by decomposing the search space to find the optimal ratio of budget $B$ over configurations $n$. Future work could investigate these applications in more detail.

Second, we presented a hybrid quantum neural network model for supervised learning. The hybrid model consisted of the combination of ResNet34 and a quantum circuit part, whose size and depth became the hyperparameters. The size and flexibility of the hybrid ML model allowed us to apply it to car image classification. The hybrid ML model with GS showed an accuracy of 0.989 after 75 iterations in our binary classification tests with images of Volkswagen Golf Hatchback 1991 and Volkswagen Beetle Hatchback 2012. This accuracy was better than of a comparable classical ML model with GS showed an accuracy of 0.920 after 75 iterations. In the same test, the hybrid ML model with TTO showed an accuracy of 0.977 after 18 iterations, whereas the comparable classical ML model with TTO, which showed the same accuracy of 0.977 after 6 iterations. Our developments provide new ways of using quantum and quantum-inspired methods in practical industry problems. In future research, exploring the sample complexity of the hybrid quantum model is of importance, in addition to generalization bounds of the quantum models similar to research in Ref.~\cite{caro2021generalization}. Future work could also entail investigating state-of-the-art improvements in hyperparameter optimization for classical and quantum-hybrid neural networks and other machine learning models by leveraging quantum-inspired or quantum-enhanced methods.

\hspace{10mm}
\section*{Declarations}

\subsection*{Competing interests}
The authors have no competing interests as defined by Springer, or other interests that might be perceived to influence the results and/or discussion reported in this paper.

\subsection*{Authors' contributions}
Andrea S., D.V.D., Alexey M. defined the research project. Asel S., A.K., Alexey M. worked on the dataset and designed classical, and quantum machine learning approaches. M.K. performed quantum neural network circuit analysis. Artem M., D.K., M.P. developed the TTO algorithm and programmed it. Asel S. and A.K. programmed and executed classical and hybrid quantum neural networks. Andrea S., D.V.D., Alexey M. analyzed the numerical data. Alexey M. supervised the research and development. All authors contributed to writing the manuscript. All authors reviewed the manuscript.

\subsection*{Funding}
Not applicable.

\subsection*{Availability of data and materials}
Dataset used to train and test the machine learning models is available in Ref.~\cite{Cars}.

\bibliography{lib}
\bibliographystyle{unsrt}
\renewcommand{\thefigure}{A\arabic{figure}}

\setcounter{figure}{0}
%\clearpage
%\newpage
\onecolumngrid
\appendix

\section*{Supplementary Material}

\section*{Quantum Circuit Analysis}\label{sec:appendix}

In this section, we critically analyze the parameterized quantum circuit (PQC) suggested in Section \ref{sec:HQNN}.  

There are many methods to do this and in this paper we focus on three of them: 

\begin{itemize}
    \item ZX calculus circuit-reducibility as suggested in Ref \cite{firstzxpaper} 
    \item Fisher information degeneracy and the effective dimension as suggested in Ref \cite{amirapaper}
    \item Fourier accessibility, first suggested in Ref \cite{schuld_fourier}
\end{itemize}

We see that the circuit in use is optimally chosen based on these measures. 

\subsection{ZX calculus}\label{sec:appendix_ZX}
\begin{figure}
    \centering
    \includegraphics[width=0.7\textwidth]{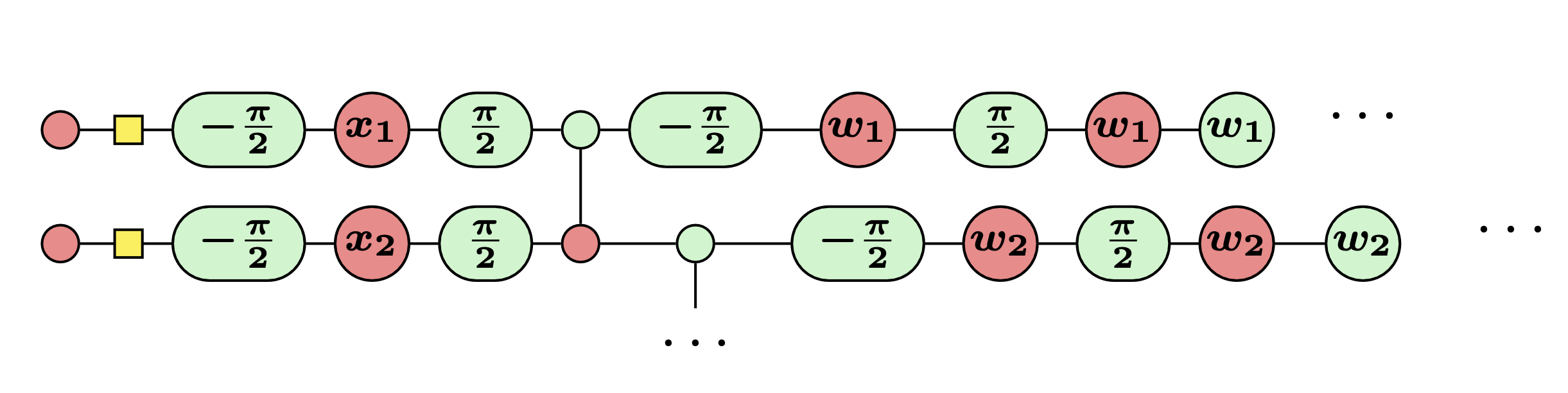}
    \caption{The only changes we could make to this circuit are fusing some constant spiders, which we will need to re-introduce later for circuit efficiency. Additionally, measurements are in the $X$-basis, so all variational parameters to the right of the last CNOT only contribute to the qubit that they are applied to. This is particularly evident in Figs \ref{fig:fourier_space}(a) and (e), where there is only one CNOT in the system.  This allows us to assign a variable specific to each qubit which we can use to tune the output of each qubit independently.}
    \label{fig:zx}
\end{figure}

ZX calculus is a graphical language that can reduce a circuit to an identical, simpler one\cite{firstzxpaper}.  To reduce a circuit using ZX calculus we need to first convert the quantum circuit to a ZX graph.  Then we can use the ZX calculus rules, suggested in Ref \cite{zx_practical}, to reduce this graph to a more fundamental version of itself. We then convert the reduced ZX graph back to a new and reduced circuit.  If a circuit cannot be reduced, we shall refer to it as ZX-irreducible. A circuit of this type can use the maximum potential of the trainable layers and includes no \emph{fully redundant parameters}. Our circuit produces the graph in Fig \ref{fig:zx}.  None of the parameterized gates shown in this figure can commute or be simplified, and therefore our circuit is ZX-irreducible. Specifically, the following two crucial steps were taken to make sure that this is the case: 

\begin{itemize}
    \item Due to the final $R_Z$ rotation gates, measurements were made in the $X$-basis to make sure these gates were not made redundant, and
    \item $R_Y$ rotation gates were employed to prevent the non-commutativity through the CNOT gates.
\end{itemize}

% \begin{figure}[b]
%     \centering
%     \resizebox{0.6\textwidth}{!}{%
%     \begin{ZX}
% \zxX*{} \ar{r} & \zxH{} \ar{r} & \zxZ{-\frac{\pi}{2}} \rar & \zxX*{x_1} \rar & \zxZ{\frac{\pi}{2}} \rar & \zxZ*{} \dar \rar & \zxZ{-\frac{\pi}{2}} \rar & \zxX*{w_1} \rar & \zxZ{\frac{\pi}{2}} \rar  & \zxX*{w_1} \rar & \zxZ*{w_1}  & \cdots\\
% \zxX*{} \ar{r} & \zxH{} \ar{r} & \zxZ{-\frac{\pi}{2}} \rar & \zxX*{x_2} \rar & \zxZ{\frac{\pi}{2}} \rar & \zxX*{} \rar &\zxZ{} \rar \dar& \zxZ{-\frac{\pi}{2}} \rar & \zxX*{w_2} \rar & \zxZ{\frac{\pi}{2}} \rar  & \zxX*{w_2} \rar & \zxZ*{w_2}  & \cdots\\
% \zxNone{}  & \zxNone{} & \zxNone{} & \zxNone{} & \zxNone{} & \zxNone{} & \cdots \\
%     \end{ZX}}
%     \caption{The only changes we could make to this circuit are fusing some constant spiders, which we will need to re-introduce later for circuit efficiency. Additionally, measurements are in the $X$-basis, so all variational parameters to the right of the last CNOT only contribute to the qubit that they are applied to.  This is particularly evident in Figs \ref{fig:fourier_space}(a) and (e), where there is only one CNOT in the system.  This allows us to assign a variable specific to each qubit which we can use to tune the output of each qubit independently.}
%     \label{fig:zx}
% \end{figure}

Although ZX-irreducibility is a crucial condition to look for, further analysis is required to understand the expressivity and the efficiency of the circuit.

\subsection{Fisher information and effective dimension}\label{sec:appendix_fim}
We can summarise supervised machine learning as creating a hypothesis model $h_{\boldsymbol{\theta}}(\hat{\mathbf{x}})$ from a labelled dataset $(\mathbf{x},y) \in \mathcal{X}\times\mathcal{Y}$ that could produce an approximation to the distribution of the data in nature, $f(\mathbf{x})$. We are provided with a subset of $S$ labelled data points from this distribution and we need to optimise our hypothesis model to be a representative model of $f(\mathbf{\hat{x}})$.  

To do this, we need to maximize the probability that given the model parameters $\boldsymbol{\theta}$ and some data point $\mathbf{x}$ we get the associated label $y$. This conditional probability can be written as $P(y|\mathbf{x},\boldsymbol{\theta})$. However, the latter notion assumes a uniform distribution over $\mathcal{X}$, and to be more accurate we need to consider the joint probability, $P(y,\mathbf{x}|\boldsymbol{\theta})$.  The joint probability distribution can be empirically evaluated for any value of $\mathbf{\theta}$ for a given subset of data. Thus, we can think of the joint probability as an N-dimensional manifold where N is the number of trainable parameters $N = |\boldsymbol{\theta}|$. The Fisher information matrix $F(\boldsymbol{\theta})$ can define a metric over this manifold\cite{amirapaper,amaripaper}
\begin{equation}\label{fishereqn}
F(\boldsymbol{\theta}) = \mathbb{E}_{\{x_i,y_i\}}[ \boldsymbol{\nabla}_{\boldsymbol{\theta}} \log{(P)}\boldsymbol{\nabla}_{\boldsymbol{\theta}}\log(P)^T  ].
\end{equation}

This metric can be diagonalized to produce a locally Euclidean tangential basis whose diagonal values provide the square of the gradient of our joint probability in this diagonalized basis. These values can be obtained by calculating the eigenvalues of the Fisher matrix. To understand the usefulness of this insight, we need to understand the issue of barren plateaus in quantum neural networks (QNNs).  Ref \cite{mcclean_2018_barren} suggested that for a chosen QNN, the expectation values of the gradients are zero and their variances decrease exponentially with the number of qubits. This combination means that QNNs suffer from vanishing gradients, a phenomenon known as the barren plateau problem. We must avoid these barren plateaus by ensuring our network can produce a spectrum of gradients rather than a large number of zeros. We showed that the eigenvalues of the Fisher information matrix produced the square of our gradients. Therefore, by calculating the eigenvalue spectrum of Fisher matrices for many realizations of $\boldsymbol{\theta}$ we can investigate the trainability - the robustness of the QNN against barren plateaus - of the specific 2-qubit network. It is noteworthy that the barren plateau phenomenon scales exponentially with the qubit count and that this section of the analysis is only applicable to the 2-qubit case.  A network with high trainability would have a lower eigenvalue degeneracy about zero. 
\begin{figure}
    \centering
    \includegraphics[width=1\textwidth]{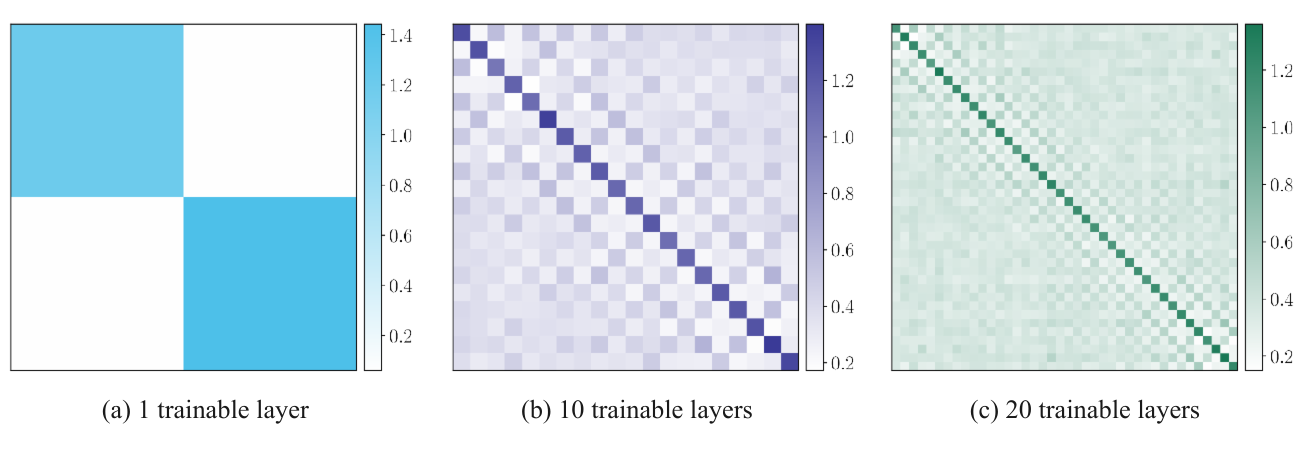}
    \caption{The square-averaged normalized Fisher matrices. The diagonals of these matrices show that the quantum circuit distributes the gradients to all trainable parameters and there is no evident single-parameter dominance.}
    \label{fig:fisher_matrix}
\end{figure}

\begin{figure}
    \centering
    \includegraphics[width=\textwidth]{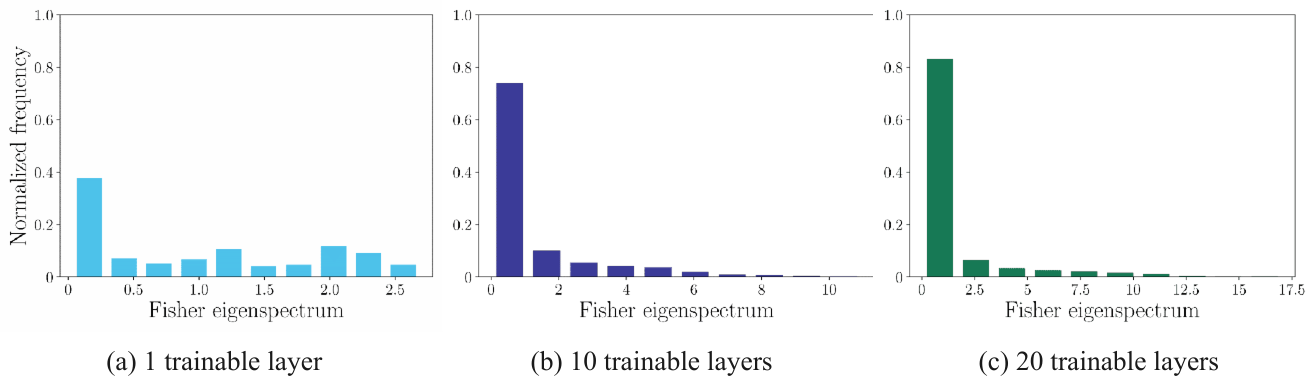}
    \caption{The normalized histogram of the Fisher eigenvalue spectra. Degeneracy about zero means that a smaller portion of the parameters is effectively driving the training, i.e. a lower trainability. (a) has the highest trainability and (c) the lowest. This means that by increasing the number of trainable layers we decrease the trainability of the model. We see in Fig \ref{fig:ed_rank} that this decrease in trainability is accompanied by an increase in model expressivity, presenting a trade-off for model selection.}
    \label{fig:ev_spectra}
\end{figure}
\begin{figure}
    \centering
    \includegraphics[width=\textwidth]{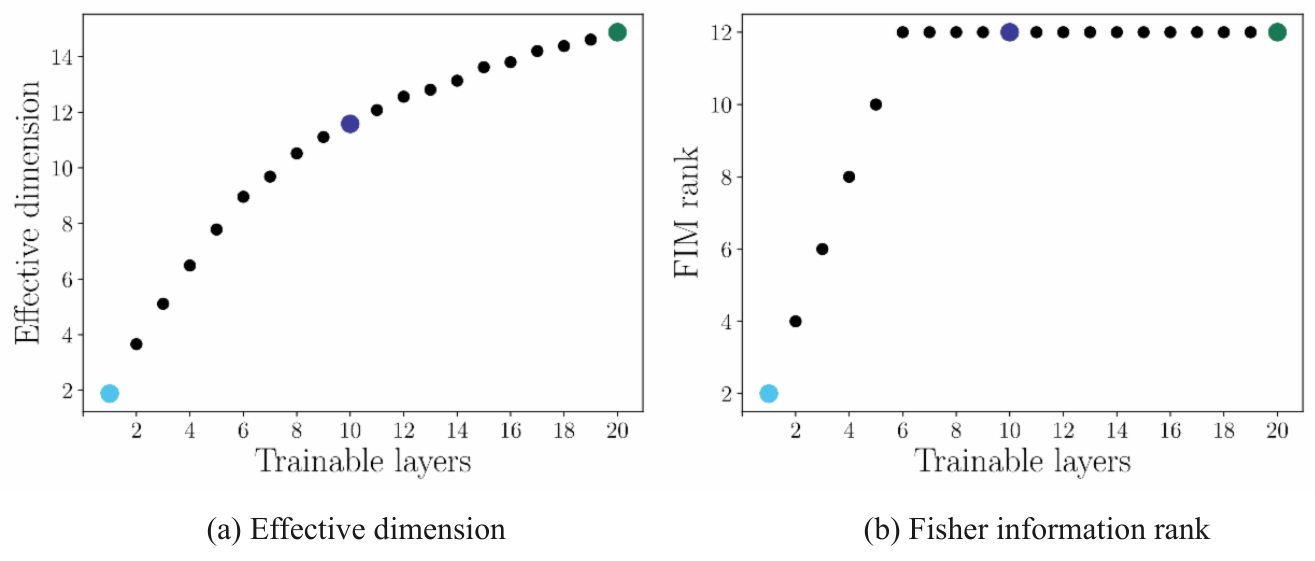}
    \caption{(a) shows that the effective dimension increases with the number of trainable layers, and (b) illustrates the limit of over-parameterization of the circuit. The colored points light blue, navy, and green correspond to the circuits with 1, 10, and 20 trainable layers, respectively. }
    \label{fig:ed_rank}
\end{figure}
Ref \cite{EDpaper} takes this concept a step further by assuming that - under some weak conditions - the Fisher information matrix above is equal\footnote{It is noteworthy that this idea is sometimes contested in the statistical learning literature \cite{thomas2020interplay,eqn35}.} to the Hessian matrix defined as 

\begin{equation}
    H(\boldsymbol{\theta}) = \mathbb{E}_{\{x_i,y_i\}}[\boldsymbol{\nabla}^T\boldsymbol{\nabla} \log(P)],
\end{equation}

which is the matrix of second-order derivatives.  Then, it uses this equivalence to derive a complexity measure that is dependent on the size of the subset $S$. This measure of complexity is defined as the effective dimension and was first practically explored in Ref \cite{amirapaper} to show that QNNs can have a higher expressivity than classical machine learning models. The latter work defines the effective dimension as
\begin{equation}
    d_{\gamma,S}(\mathcal{M}_\Theta):=2\frac{\text{log}\left(\frac{1}{V_\Theta}\int_\Theta\sqrt{\text{det}\left(id_N+\frac{\gamma S}{2\pi \text{log} S}\hat{F}(\boldsymbol{\theta})\right)}d\boldsymbol{\theta}\right)}{\text{log}\left(\frac{\gamma S}{2\pi \text{log} S}\right)},
    \label{EDEqn}
\end{equation}
where $V_\Theta := \int_\Theta d\theta $ is the volume of the parameter space, $\gamma$ is a constant in $(0,1]$ introduced in Ref \cite{amirapaper} , and $\hat{F}(\boldsymbol{\theta})$ is the normalised Fisher matrix defined as
\begin{equation}
\label{normalisedfisher}
    \hat{F}_{ij}(\boldsymbol{\theta}):=N\frac{V_\Theta}{\int_\Theta tr(F(\boldsymbol{\theta}))d\boldsymbol{\theta} } F_{ij}(\boldsymbol{\theta}).
\end{equation}

%% add some results here ---------------------------------------------------

We can calculate the Fisher information for the specific hyperparameter settings of our circuit.  Specifically, we consider a 2-qubit variation of this circuit with the number of trainable layers varying from $1$ to $20$.  Following the lead of Ref \cite{amirapaper}, we create a Gaussian dataset $\mathbf{x} \sim \mathcal{N} (\mu = 0, \sigma^2 = 1)$ and evaluate the joint probability by overlapping the specific resultant state with the state of our QNN
\begin{equation}
    P(y,\mathbf{x}|\boldsymbol{\theta}) = \bra{~y~}\ket{~\psi(\boldsymbol{\theta},\mathbf{x})~},
\end{equation}
where $y$ is the output state. Note that this has to be averaged over all possible $y$ and $\mathbf{x}$. This way, we can calculate the empirical Fisher information for any $\boldsymbol{\theta}$.  Fig \ref{fig:fisher_matrix} shows the mean-square, normalised Fisher matrix for $1000$ data points and $100$ uniform weight realizations $\theta \in (0,2\pi]$. Observing the diagonal elements, it seems that none of the parameters is especially dominant or redundant. A further test would be to look at the Fisher eigenvalue spectra shown in Fig \ref{fig:ev_spectra}. We can see that the degeneracy of the eigenvalues around zero increases for a higher number of trainable layers. 

Finally, to obtain the effective dimension, we can evaluate the integral in Eq \ref{EDEqn} by taking the average of $100$ Fisher realizations. Fig \ref{fig:ed_rank}(a) shows the effective dimension against the number of trainable layers of our network.  Increasing the number of trainable layers increases the effective dimension.  This is unsurprising as we defined the effective dimension as a measure of expressivity and we expect that adding trainable layers would increase the expressivity of the network. However, we also see that adding trainable layers could yield diminishing returns at higher values. 

Additionally, it was shown in Ref \cite{overparam} that certain QNNs can become over-parameterized and exhibit lowered parameter efficiency.  This was quantified by finding the parameterization for which, at least at one point in the loss landscape, any added parameter would leave the rank of the Fisher information matrix unchanged - in other words, the rank of the Fisher matrix becomes saturated for an over-parameterized circuit. Examining Fig \ref{fig:ed_rank}(b), we see the FIM rank of the circuit increases linearly with the number of trainable parameters and then plateaus at 6 trainable layers, reaching a maximal rank of $r=12$. This means that although the effective dimension seems to increase beyond this point, but the circuit is saturated and there is no further increase in expressivity. 

These analyses signify a trade-off between trainability, determined by the eigenvalue spectra in Fig \ref{fig:ev_spectra}, and the expressivity quantified by the effective dimension and upper-bound by the maximal rank. 
%% end of results -----------------------------------------------------------
\subsection{Fourier accessibility}\label{sec:appendix_fourier}

\begin{figure}
    \centering
    \includegraphics[width=\textwidth]{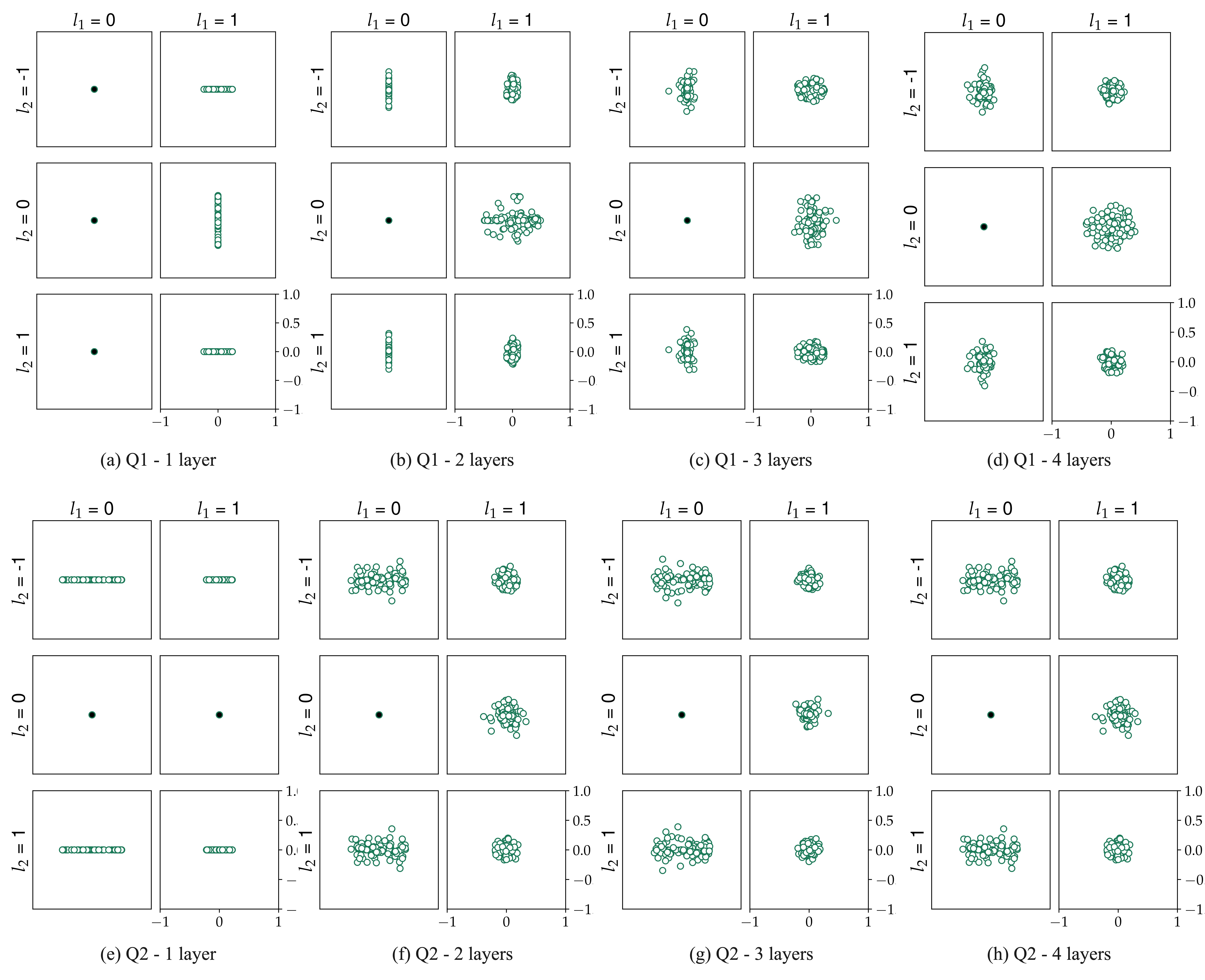}
    \caption{The Fourier accessibility of our circuit when employed for two features.  Note that $l_1$ is only shown to span the positive half of its indices and this is due to the symmetry of these coefficients $c_{l_1,l_2} = c_{-l_1,-l_2}$.  In all figures, we see that the circuit is not able to change the offset coefficient.  This sets a constraint but also increases our accessibility because at a maximum the amplitude of the Fourier output cannot exceed 1 (as the expectation value of our circuit needs to remain between -1 and 1). Figs (a) and (e) respectively show the output of the first and the second qubits when only one trainable layer is implemented. In this case, the associated qubit has an asymmetric advantage in accessibility.  We also see that in this case, the phases of our coefficients remain fixed.  Both of the mentioned issues can be improved by adding trainable layers, and Figs (d) and (h) corroborate this statement. Note that this does not show the full extent of Fourier accessibility as a complete investigation would also look into the inter-dependence of these coefficients, but that lies outside the scope of this work.}
    \label{fig:fourier_space}
\end{figure}

Ref \cite{schuld_fourier} showed that a QNN that uses angle-embedding\footnote{Ref \cite{schuldkernel} provided expressivity analysis for different angle embedding strategies.} produces a truncated Fourier series of degree $L$.  This degree is dependent on the number of encoding repetitions employed in a QNN, a strategy first employed in  Ref \cite{data_reuploading}. Furthermore, Refs \cite{schuld_fourier,qaum} showed that for a multi-feature setting we get a multi-dimensional truncated Fourier series. For a two-feature setting, we get 
\begin{eqnarray}
    f(\boldsymbol{\theta},\boldsymbol{x}) = \bra{\psi(\boldsymbol{\theta},\boldsymbol{x})}M\ket{\psi(\boldsymbol{\theta},\boldsymbol{x})}\\
    f(\boldsymbol{\theta},\boldsymbol{x}) = \sum_{l_1=-L_1}^{L_1}\sum_{l_2=-L_2}^{L_2} 2|c_{l_1,l_2}| cos(l_1 x_1 + l_2 x_2 - arg(c_{l_1,l_2})),
\end{eqnarray}
where $\ket{\psi(\boldsymbol{\theta},\boldsymbol{x})}$ is the quantum state of the system after encoding and variational layers, $M$ is the measurement gate, and $L_1$ and $L_2$ are the number of encoding repetitions of the first and the second feature respectively.  The complex coefficients $c_{l_1,l_2}$ determine the amplitude and the phase of each Fourier term. These coefficient depend only on the variational gates, and so, our accessibility to a full Fourier series is limited by how these variational gates span the Fourier space.    We can investigate a specific subset of our networks with 2 features and a single encoding repetition.  This means that our circuit has $L_1 = L_2 = 1$.  Thus, we can set up the circuit and randomly realise the weights many times to assess the Fourier accessibility of the circuit. Fig \ref{fig:fourier_space} shows the Fourier accessibility of our network for 100 uniform realizations of weights $\boldsymbol{\theta} \in [0,2\pi)^N$.  It is evident that increasing the number of trainable layers improves the Fourier accessibility of the QNN. Furthermore, we can see that to have an unimpeded network we need at least 3 layers of variational gates.

\subsection{Summary}

In this analysis, we assessed the feasibility of the chosen quantum circuit and looked at three approaches for analyzing its effectiveness: ZX-reducibility,  Fisher information, and Fourier analysis.  In Sec \ref{sec:appendix_ZX} we proved that there are no redundant parameters in the circuit caused by commutation of the quantum gates and that certain weights are reserved for independent contribution to each qubit.  Then, in Sec \ref{sec:appendix_fim} we showed that none of the parameters dominated the training and that by increasing the number of trainable layers, the trainability and the complexity of our model respectively decreased and increased. The increase in model complexity stopped at 6 layers, where we showed that the rank of the Fisher information matrix was saturated and any additional parameterization would be futile.  Finally, in Sec \ref{sec:appendix_fourier} we used the theoretical findings of Ref \cite{schuld_fourier} to show that for a 2-qubit version of our network we at least needed 3 layers of variational gates to represent the full Fourier landscape.  

\end{document}